\documentclass[a4paper,11pt]{article}
\pdfoutput=1 

\usepackage{jcappub} 

\usepackage{amsmath}
\usepackage{enumerate}
\usepackage{graphicx}
\usepackage{caption}
\usepackage{subcaption}
\usepackage{color}
\usepackage{float}
\usepackage{pdfpages}

\def\f{\frac}

\numberwithin{equation}{section}
\begin{document}

\author[a]{George F.R. Ellis}
\emailAdd{george.ellis@uct.ac.za}
\affiliation[a]{Astrophysics, Cosmology and Gravity Centre, Department of Mathematics and Applied Mathematics, University of Cape Town, Rondebosch 7701, Cape Town, South Africa}

\author[a]{Emma Platts}
\emailAdd{pltemm002@myuct.ac.za}

\author[b]{David Sloan}
\emailAdd{david.sloan@physics.ox.ac.uk}
\affiliation[b]{Beecroft Institute of Particle Astrophysics and Cosmology, Department of Physics,
University of Oxford, Denys Wilkinson Building, 1 Keble Road, Oxford, OX1 3RH, UK}

\author[a]{Amanda Weltman}
\emailAdd{amanda.weltman@uct.ac.za}

\title{Current observations with a decaying cosmological constant allow for chaotic cyclic cosmology}

\abstract{We use the phase plane analysis technique of Madsen and Ellis \cite{Madsen:1992tv} to consider a universe with a true cosmological constant as well as a cosmological ``constant" that is decaying. Time symmetric dynamics for the inflationary era allows eternally bouncing models to occur. Allowing for scalar field dynamic evolution, we find that if dark energy decays in the future, chaotic cyclic universes exist provided the spatial curvature is positive. This is particularly interesting in light of current observations which do not yet rule out either closed universes or possible evolution of the cosmological constant. We present only a proof of principle, with no definite claim on the physical mechanism required for the present dark energy to decay.} 

\arxivnumber{1511.03076}

\maketitle
\flushbottom

\section{Introduction}
\subsection{Overview}
 
Cosmology, being the study of the large scale universe, confronts questions relevant to the very core of existence. As a serious scientific field for only a century, the field of cosmology has advanced rapidly to become a precision science. The age old belief in a static universe has been dislodged by a confrontation with data which tells us that the universe is not only evolving and expanding, but presently accelerating in its expansion \cite{Riess:1998cb, Eisenstein:2005su, Dunkley:2008mk, Leibundgut:1996qm}. However, questions about the very beginning of the universe remain unresolved given all available data. On the largest scales, the universe appears isotropic and homogeneous, respecting the cosmological principle. The observations point to a dynamic universe that has evolved through different phases, each characterised by the domination of a different matter type.  The best explanation to date is given by a phase of inflation \cite{Starobinsky:1980te, Guth:1981uk, Albrecht:1982wi, Linde:1981mu} in the early universe, lasting at least 60 e-folds, followed successively by a radiation phase, a matter dominated phase, and a dark energy dominated phase at the present time. Although we typically assume a singularity prior to inflation, a variety of proposals for cyclic models \cite{Steinhardt:2001st, Khoury:2001wf, Khoury:2003rt, Turok:2004yx,Ashtekar:2006wn} are not yet ruled out, though none have yet provided a satisfactory bounce mechanism together with a model of a transition from the current era of accelerated expansion to a collapsing phase without invoking new physics \footnote{Penrose' conformal cyclic cosmology \cite{Penrose:2011zz} bypasses this latter requirement at the cost of replacing general relativity by some kind of conformal generalisation. Loop quantum cosmology models require new physics.}.  \\

In this paper, we will show that a cyclical universe is reasonable from a dynamical systems perspective, and requires in addition to standard cosmological assumptions, only two conditions; (i) the spatial sections must have positive spatial curvature ($\Omega_k = +1$), and (ii) the late time effective cosmological ``constant" must decay fast enough as a function of the scale factor. Both of these conditions are consistent with all current observations to date. For the equation of state we have  \cite{Ade:2015xua,Kumar:2012gr}
\begin{equation}\label{eq:de}
w:=\frac{p}{\rho} = -1.006^{+0.085}_{-0.091}\,,\,\, \frac{dw}{da}=0.03^{+0.68}_{-0.40}. 
\end{equation} 
For the curvature we have  \cite{Wang:2007mza}
\begin{equation}\label{eq:om1}
\Omega_k=0.006^{+0.012+0.032}_{-0.013-0.025}
\end{equation} with constant equation of state $w(a) = w_0$ and 
\begin{equation}\label{eq:om2}
\Omega_k=0.002^{+0.018+0.032}_{-0.018-0.041}
\end{equation} for $w(a)=w_0+w_a(1-a)$ \footnote{Note that we are using the convention where $\Omega_K$ is the same sign as $k$ and have translated the observational constraints accordingly. Hence \{$\Omega_K > 0 \Leftrightarrow k > $\} is positive curvature, corresponding to a 3-sphere.}.  
It is worth pointing out that the constraints quoted above are very much model dependent and for some models of dark energy, curvature and dark energy are strongly degenerate and neither is very well constrained. Using a phase plane analysis to explore the evolution of the density parameter $\Omega$ under different conditions, we will show these conditions are compatible with chaotic cyclic models, with no more exotic physics than that the dark energy decays away at late times. Our model is somewhat similar to one proposed by Starobinsky \cite{Starobinksy}, of which we were made aware after completing this work. However, we deal with a far more general case than that of a simple scalar field, and give a more complete analysis of the overall behaviour of these space-times in terms of phase planes.

\subsection{Chronology of the Universe}
 
We initially model the chronology of the universe in accordance with the prevailing standard scientific view - a Big Bang followed by inflation \cite{Lemaitre:1927zz,Hubble:1929ig,Starobinsky:1980te,Guth:1981uk,Albrecht:1982wi,Linde:1981mu}. After the initial Big Bang singularity and before inflation, little is known about the universe. The universe was so hot (at temperatures of $\sim10^{15}$K) that no particles could exist with any sort of stability, and the four fundamental forces are believed to have been merged as one ``unified force''. It may be that a period of radiation preceded inflation; though this is not required for successful inflation, we will include this possibility to consider the effects on our dynamics. During cosmological inflation the universe increased in size by a factor of at least 10$^{27}$ \cite{Starobinsky:1980te,Guth:1981uk,Albrecht:1982wi,Linde:1981mu}. After inflation, the universe returned to radiation domination until it cooled to a point that allowed the forces to undergo symmetry breaking, eventually causing the separation of the strong force from the electroweak force. This meant that the first particles could finally exist in a quark-gluon plasma. The universe continued to cool until the four forces took their present forms, and the spectacular range of particles we know today could become possible. This led to the universe, at last, being matter dominated, with the densities of non-relativistic matter and radiation equal. Recombination then took place, with the emergence of the first neutral atoms. By the end of recombination, most of the protons were bound in these neutral atoms, allowing for previously scattered photons to escape freely into space, making the universe transparent. This is known as decoupling, with these photons still detectable today in the cosmic microwave background. This CMB was detected in 1965 by Penzias and Wilson - a landmark test of the Big Bang model \cite{Penzias:1965wn}. With the fundamental particles and forces in place, the universe continued to expand, giving rise through cosmic evolution to the first stable large-scale structures, such as stars, quasars and galaxies. As for the future of our universe, there are several different theories.\\

Stars will eventually die, and fewer will be born to replace them. This will lead to a darkening universe, and the eventual end of the Stelliferous Era. After this, there are a number of possible outcomes. Initially, before the discovery of the accelerating expansion of the universe, there were two main competing models - that of the ``big crunch'' and the ``big freeze'' or heat death. The big crunch assumes that the overall density of the universe is large enough such that it will eventually cease to expand and instead begin to contract. This symmetric theory then sees the collapse of space-time into a singularity. Heat death, on the other hand, holds the view that the universe will continue to expand and gradually approach absolute zero temperature. Eventually, in a state of maximum entropy, the universe will no longer be able to sustain life, leaving the universe to disintegrate into empty space and weak radiation, infinitesimally close to absolute zero temperature. \\

In order to decide the most likely model, astronomers in the early 1990's calculated the total mass density of the universe. The low density detected lead them to believe that heat death is the most likely scenario. It was not long after however, with the initial discovery in 1998, that the universe was found to be expanding at an accelerating rate \cite{Mukherjee:2003rc,Perlmutter:2003kf,Tonry:2003zg}. The observations of the Type Ia supernovae confirm this, with the widely accepted belief that dark energy is accountable \cite{Leibundgut:1996qm}. With dark energy, heat death is confirmed as being be the likely fate of our universe, with the universe expanding at an ever increasing rate and all matter and radiation fading away.  \\

However an alternative to the ever-expanding universe is a cyclic universe, which undergoes successive cycles of expansion and collapse.  Such models have been widely explored in the past century, see for example \cite{Tolman1934,Dicke:1900mn,Steinhardt:2001st, Steinhardt:2001vw, Mukherji:2002ft, Khoury:2001wf, Khoury:2003rt, Turok:2004yx, Penrose:2008, Penrose:2009, Gurzadyan:2010da,Penrose:2011zz, Sahni:2012er, Sahni:2015kga}. However none of them are fully satisfactory. This paper makes a proposal for cyclic models that is very conservative in physical terms: its essential physical requirement is only dark energy that dies away as the universe expands.  

\subsection{Modelling the currently accepted chronology}
 
The previously described chronology can be divided into five epochs (radiation, inflation, radiation, matter and dark energy dominated).   In this paper, these are developed following the phase plane method of Madsen and Ellis \cite{Madsen:1992tv}, with addition of a final phase taking into account the cosmological constant, $\Lambda$. The case where $\Lambda$ decays is then also explored. We note that current inflationary literature often no longer includes a radiation phase prior to the start of inflation, however in keeping with Madsen and Ellis \cite{Madsen:1992tv}, we will include this phase at the outset. Whether it is included or not makes no difference to the cyclic models; however its inclusion allows a simple representation also of emergent universe models \cite{EllMaa04,Ellis:2003qz,Guendelman:2015qta}.  

\section{Phase Planes}
\label{phase}
\subsection{Phase plane equations}
\label{phaseeq}
 
Assuming an FLRW universe described by a scale factor $a$, Hubble parameter $H \equiv\dot{a}/a$, and the energy-momentum tensor for a perfect fluid, $T^\mu_\nu=$diag($\rho, -p, -p, -p$), the Friedmann, Raychaudhuri, and conservation equations are, respectively, 
\begin{equation}
\label{freidone}
H^2 = \frac{8\pi G \rho}{3} - K,
\end{equation}
\begin{equation}
\label{freidtwo}
3\dot{H} + 3H^2 + \frac{8\pi\emph{G}}{2}\Big(\rho + 3p\Big) = 0,
\end{equation}
\begin{equation}
\label{cons}
\dot{\rho} + 3H(\rho + p) = 0.
\end{equation}
 
Here $\rho$ is the total energy density, $p$ is the total pressure and the curvature, $K=k/a^2$, is set by the value of the constant $k$, i.e. $k>0$ for a closed universe, $k=0$ for a flat universe and $k<0$ for an open universe.\\

Defining the equation of state $w\equiv-p/\rho$, we note that it varies with epoch. Since the scale factor $a(t)$ determines conditions at a time $t$, the equation of state is written $w=w(a)$. Initially, we will treat each epoch as having a constant equation of state as determined by the dominant matter type. 
With $\kappa\equiv8\pi G$, the total density parameter $\Omega$ is defined as 
\begin{equation}\label{eq:Omega}
\Omega\equiv\frac{\kappa\rho}{3H^2}.
\end{equation} 
Inserting this into Equation \ref{freidone}, we arrive at
\begin{equation}
\label{om}
K = H^2(\Omega - 1).
\end{equation}
This can be written in terms of the spatial curvature density $\Omega_K=\frac{K}{H^2}$ as
\begin{equation}
\Omega_K=\Omega-1.
\end{equation}
 
By examining the above equation we can see that $\Omega_K<0$ if $\Omega<1$, $\Omega_K=0$ if $\Omega = 1$ and $\Omega_K> 0$ if $\Omega>1$; the cases for negative, flat and positive spatial curvature respectively. If the spatial sections have their simply connected natural topology, these are also the cases of  open, flat and closed universes respectively. \\

Using the conservation equation (Equation \ref{cons}) and Raychaudhuri equation (\ref{freidtwo}), we find the derivative of the density parameter $\Omega$ to be given by
\begin{equation}
\label{phaseone}
\dot{\Omega} = -H\Omega(1 - \Omega)(1 + 3w).
\end{equation}
 
Note that $\Omega = 0$ (an empty model) and $\Omega = 1$ (flat spatial sections) are solutions with constant $\Omega$ no matter what the equation of state. Also $w = -1/3$ and $H=0$ imply $\Omega$ is constant, whatever its value. 

\subsection{Phase planes for epochs with a constant equation of state}
\label{sectphase}
 
In order to create the phase planes, we consider the situation in which the equation of state is constant in each epoch, each of which is defined by initial and final values of the Robertson-Walker scale factor. Each epoch corresponds to when the universe was dominated by a particular simple, one-component field. 

\subsubsection{Standard matter case: $w>-1/3$}
 
\begin{figure}[!ht]
\centering
\begin{subfigure}{0.4\textwidth}
\includegraphics[trim = 1cm 9cm 3cm 9cm,width=0.9\linewidth]{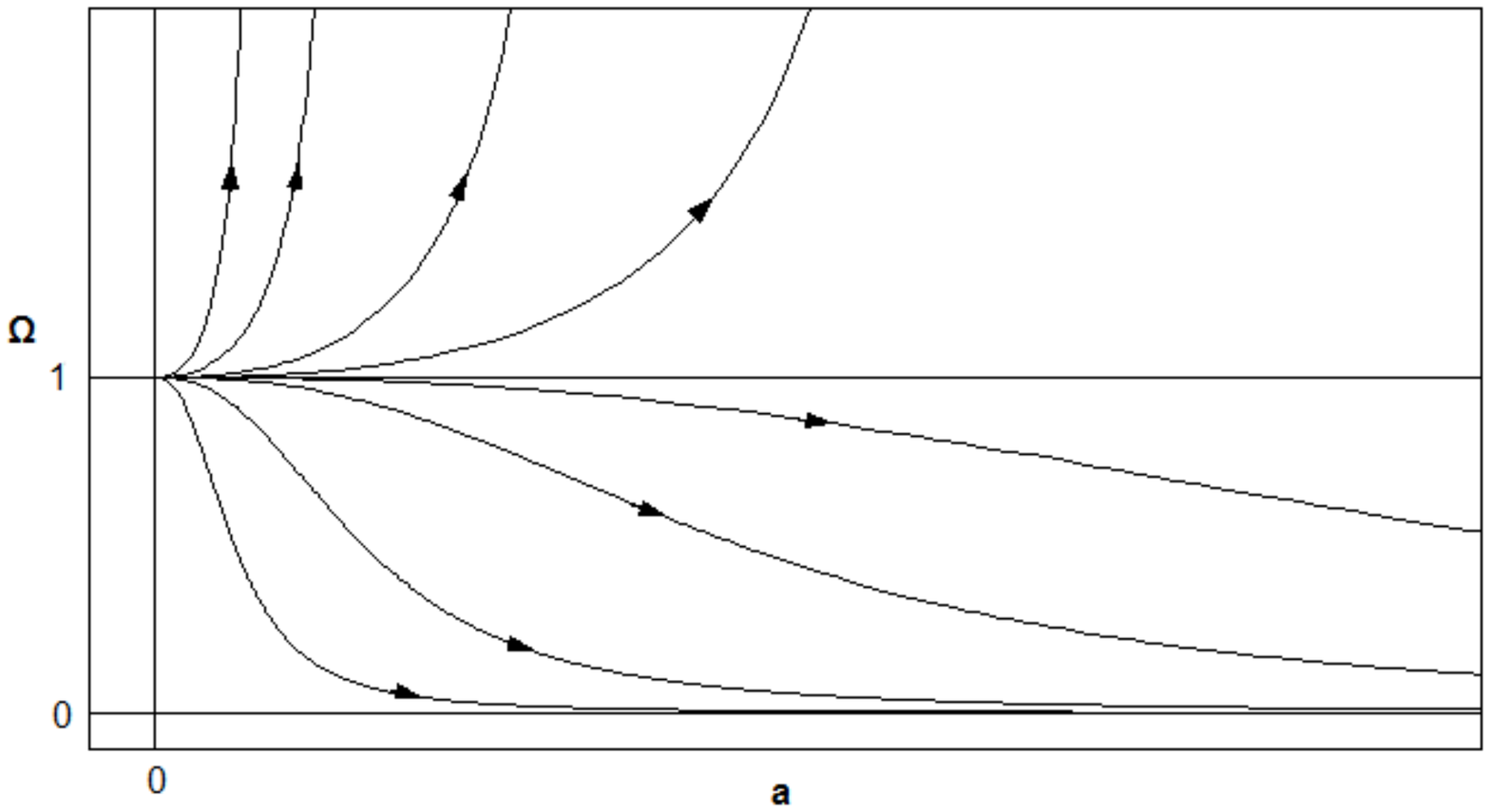}
\caption{$w=0$ \label{rad}}
\end{subfigure}
\begin{subfigure}{0.4\textwidth}
\includegraphics[trim = 1cm 9cm 3cm 9cm,width=0.85\linewidth]{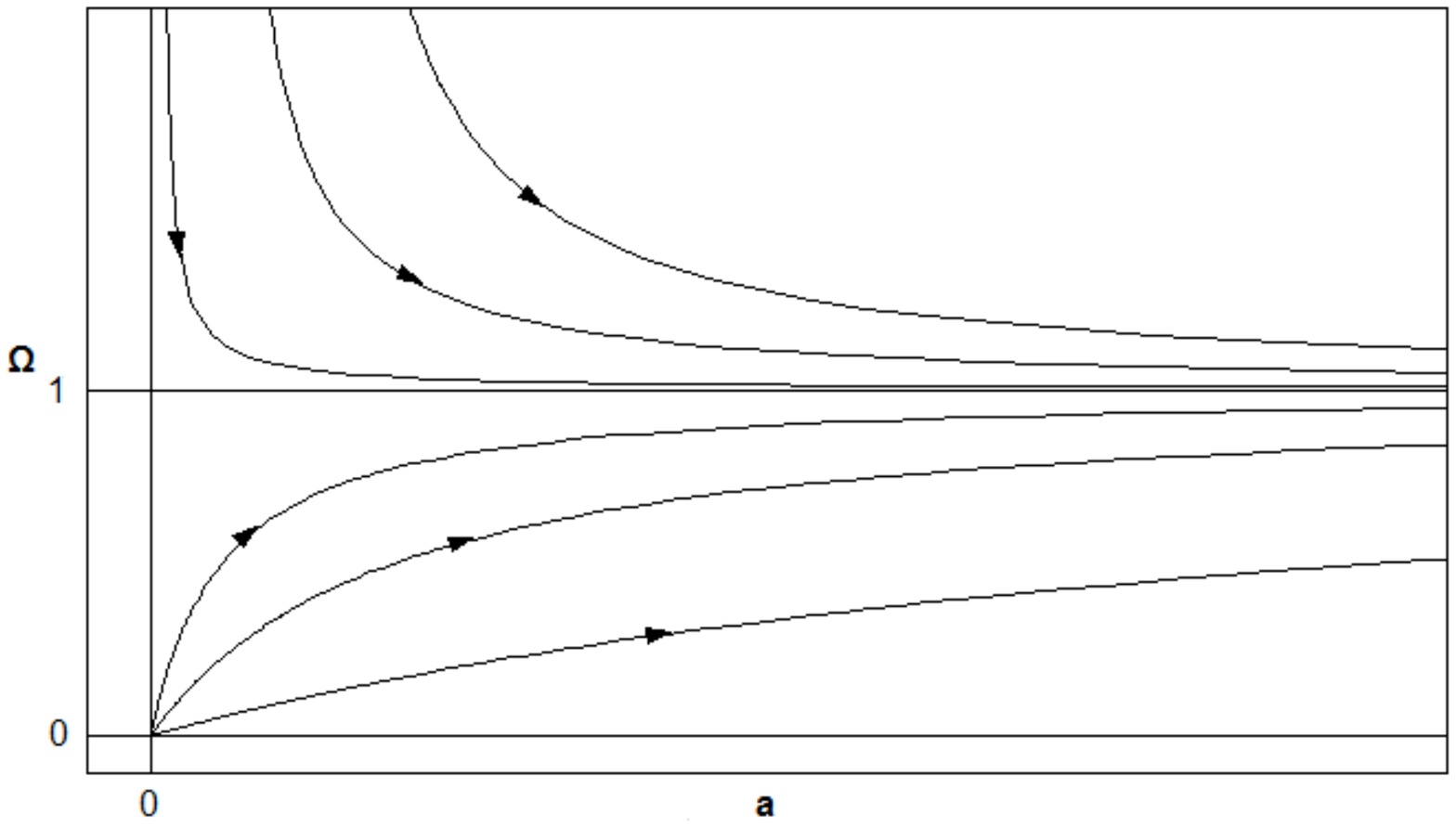}
\caption{$w=-1$ \label{inf}}
\end{subfigure}
\caption{Phase planes for dust $w=0$ and slow roll inflation $w=-1$, respectively}
\end{figure}

Standard matter has $w \geq 0$, but the phase diagram stays the same provided $w > -1/3$. Important examples are $w=0$ and $w=1/3$ for pressure-free matter (dust) and pure radiation, respectively. 
In FIG. \ref{rad} we can see that $\Omega = 1$ is always an unstable asymptote; the curves cannot cross this line. For $\Omega>1$ the curves increase monotonically, diverging to infinity for finite values of $a$ (where a maximum occurs as $H \rightarrow0$). For $\Omega<1$, the curves decrease monotonically towards $\Omega = 0$ as $a \rightarrow \infty$. 

\subsection{Inflationary case: $-1\leq w< -1/3$}
 
Inflation requires a phase of accelerated expansion with an equation of state with negative pressure such that $w < -1/3$; physical plausibility requires $w \geq -1$. In order to agree with observations of the spectral index, we require an equation of state $w\approx-1$ \cite{Ade:2015lrj}. In FIG. \ref{inf}, all curves monotonically tend to the stable asymptote $\Omega=1$, so they tend to spatial flatness. They cannot cross this line: hence should the universe have been open prior to inflation ($k = -1 \Leftrightarrow \Omega < 1$, see equation (\ref{om})), the universe necessarily stays open. This can be seen by the way in which phase plane lines do not cross the line $\Omega=1$. 

\section{A Cosmological Constant}
\label{sectcos}
 
We next seek to include the current late time phase of accelerated expansion of the universe, usually attributed to dark energy \cite{Mukherjee:2003rc,Perlmutter:2003kf,Tonry:2003zg}.

\subsection{The phase plane equations with dark energy} 
 
Dark energy theories include a simple a cosmological constant or some evolving field, see for example \cite{Caldwell:1998je,Steinhardt:2003st, Khoury:2003rn,Khoury:2003aq,Brax:2004qh,Brax:2004px}. With a cosmological constant, $\Lambda$, the Friedmann, Raychaudhuri and conservation equations (\ref{freidone}), (\ref{freidtwo}), (\ref{cons})  become
\begin{equation}
\label{cosmoone}
{H^2 = \frac{\kappa\rho}{3} - K + \frac{\Lambda}{3}},
\end{equation}
\begin{equation}
\label{cosmotwo}
{\dot{H} + H^2 + \frac{\kappa}{6}\Big(\rho + 3p\Big) - \frac{\Lambda}{3} = 0,}
\end{equation} 
\begin{equation}
\dot{\rho}=-3H\rho(1+w)-\frac{\dot{\Lambda}}{\kappa}.
\label{rhodot}
\end{equation}
 
We can represent a cosmological constant either as a fluid with density $\rho$ and pressure $p = -\rho$, or by the term  $\Lambda$. We will follow the latter course. Defining
\begin{equation}
\Omega_\Lambda = \frac{\Lambda}{3H^2},
\end{equation}
we find the modified total density parameter \ref{eq:Omega},
\begin{equation}
\Omega=\frac{\kappa\rho}{3H^2}+\frac{\Lambda}{3H^2}.
\end{equation}
Inserting this into \ref{cosmoone}, we again obtain
\begin{equation}
K = H^2(\Omega - 1),
\end{equation}
and hence
\begin{equation}
\Omega_K=\Omega-1.
\end{equation}
 
The consequent modified equation \ref{phaseone} for $\dot{\Omega}$ is
\begin{equation}
\dot{\Omega} = -H\left(\Omega-\Omega_\Lambda\right)\left(1-\Omega\right)\left(1+3w-\frac{2\Omega_\Lambda}{\Omega-\Omega_\Lambda}\right).
\label{omcos}
\end{equation}
In the rest of this section, we consider the case of a cosmological constant ($\dot{\Lambda}=0$); in the next section, we consider the case of dynamic dark energy  ($\dot{\Lambda}\neq 0$).

\subsection{{\normalsize A pure cosmological constant:} $\bf{\Omega=\Omega_\Lambda}$}
 
We first consider the simplest case of a cosmological constant, $\Lambda$, driving the acceleration. For a pure cosmological constant dominated epoch, $\rho = 0$, $\dot{\Lambda}=0$ in the above equations, to yield FIG. $\ref{omegalam}$. For all values of $\Omega$ the universe expands, with the ultimate fate corresponding to a de Sitter exponentially expanding end state. It will have positive spatial curvature if $\Omega >1$, and negative spatial curvature if $\Omega < 1$.

\subsection{Matter plus a cosmological constant: 
	 $\bf{\Omega=\Omega_m+\Omega_\Lambda}$}
 
\begin{figure}[!ht]
\centering
\begin{subfigure}{0.4\textwidth}
\includegraphics[trim = 4cm 9cm 4cm 9cm,width=0.75\linewidth]{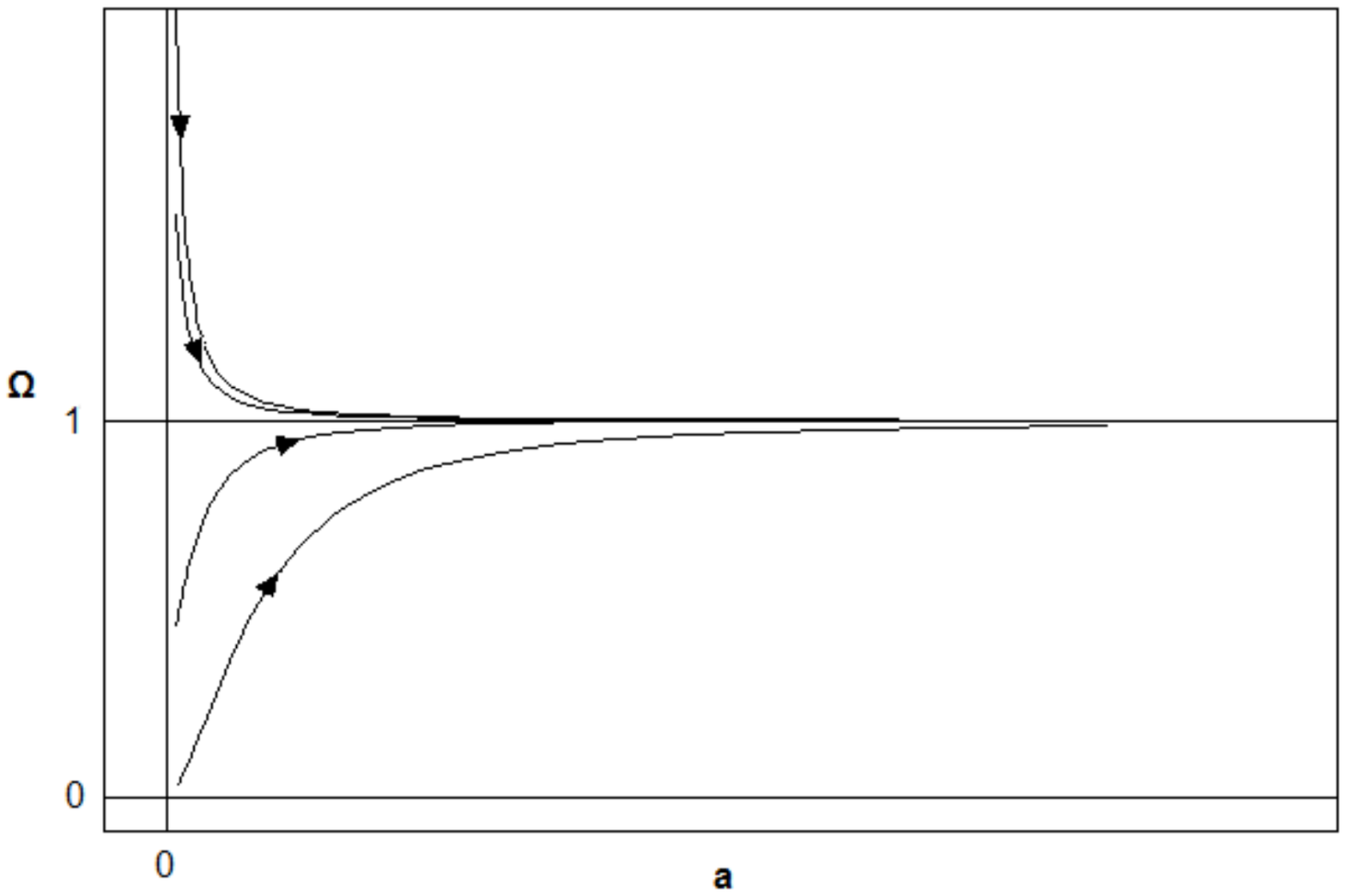}
\caption{$\Omega=\Omega_\Lambda$ \label{omegalam}}
\end{subfigure}
\begin{subfigure}{0.4\textwidth}
\includegraphics[trim = 2cm 9cm 4cm 9cm,width=0.85\linewidth]{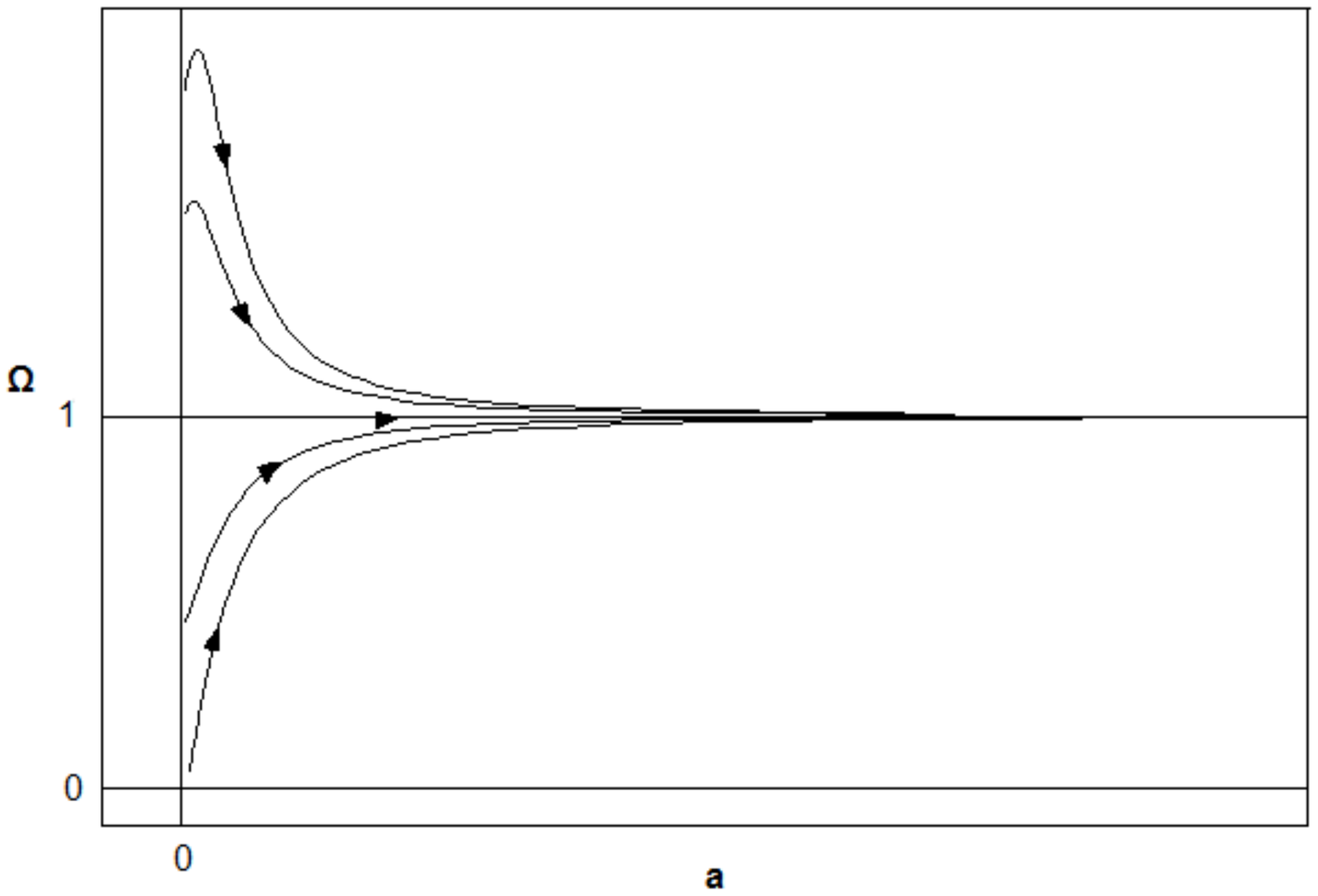}
\caption{$\Omega=\Omega_m+\Omega_\Lambda$ \label{omlammat}}
\end{subfigure}
\caption{Phase planes for pure cosmological constant, and for matter and a cosmological constant, respectively.}
\end{figure}

Given current observations, it is perhaps more pertinent to consider a universe made up of both matter and dark energy. Solving \ref{omcos} for the case of a cosmological constant ($\dot{\Lambda} = 0$) and pressure-free matter ($w = 0$), the results are presented in FIG. \ref{omlammat}, with matter causing the density to initially increase for $\Omega>1$, but the curve $\Omega=1$ is still the final asymptote. 

\section{All the phases}
\label{sec:All} 
 
Next, we put together the phase portraits for each epoch to create a phase plane for the entire history of a universe with a late time cosmological constant. \\

\subsection{A true cosmological constant}
 
\begin{figure}[h] 
\centering
	\includegraphics[trim = 0cm 0cm 0cm 0cm,scale=0.6]{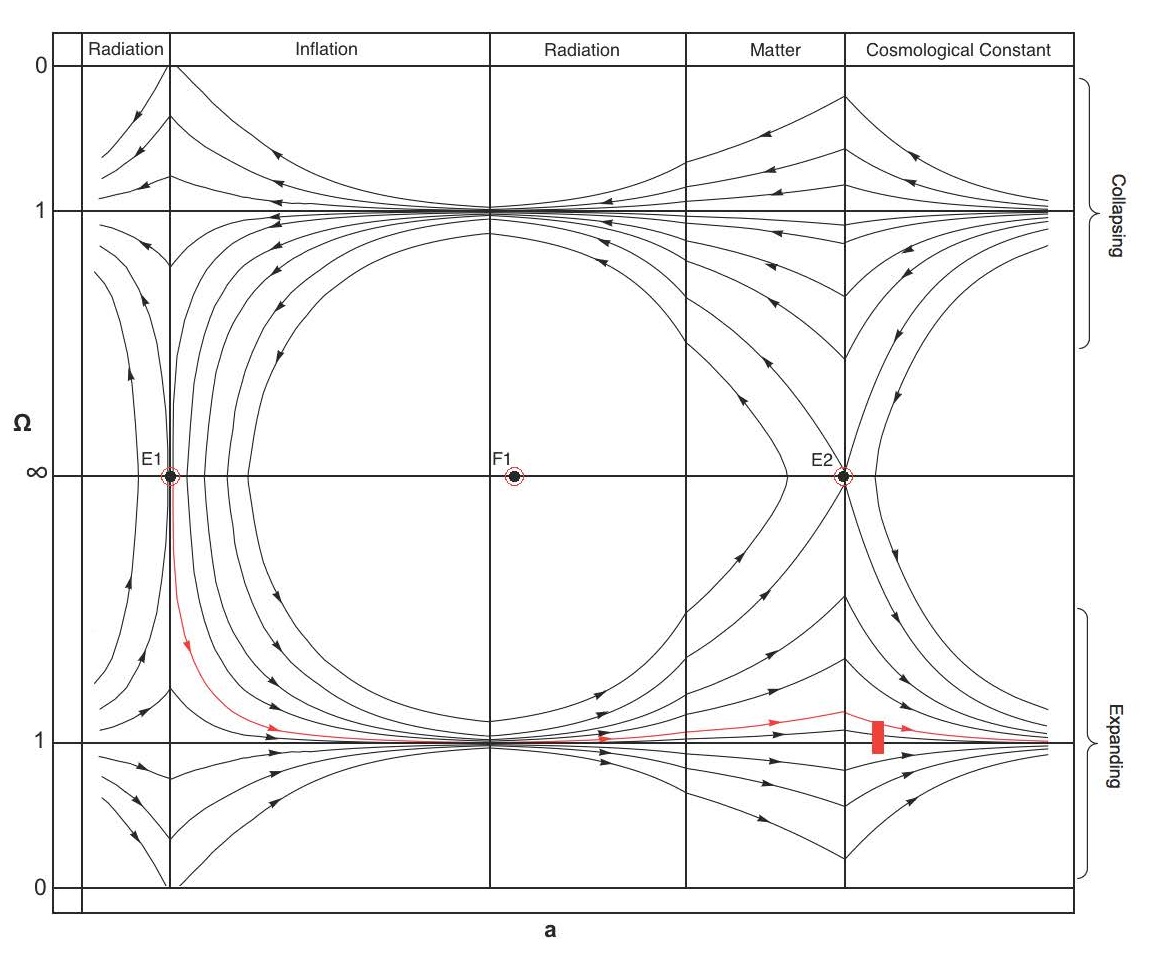}
	\caption{Time symmetric phase plane with a true cosmological constant}
	\label{fullplot}
\end{figure}

There is an unknown (presumably quantum gravity) epoch on the far left, which we do not model. We represent five phases, each separately given by the dynamical equations above with suitable choices for the matter, namely:
\begin{enumerate}
	\item A radiation dominated phase before inflation starts,
	\item The inflationary epoch,
	\item A radiation dominated Hot Big Bang epoch,
	\item A matter dominated epoch,
	\item A late time cosmological constant dominated epoch.  
\end{enumerate}
They are joined together by suitable junction conditions to give the whole diagram in FIG. \ref{fullplot},  which updates the image as found in Madsen and Ellis \cite{Madsen:1992tv} to include a cosmological constant.  Note that the scale for $a$ is non-linear. \\
 
We assume a time-symmetry between the collapsing and expanding phases, as in Madsen and Ellis \cite{Madsen:1992tv}. It is thus of interest to transform the infinities of $\Omega$ (created because $H=0$ at a point of maximum expansion) to finite values, using the variable change $\omega=$arctan(log($\Omega$)) to create a time-symmetric phase plane for states of expansion and collapse, joined at the boundary $\Omega = \infty$ as shown in FIG. \ref{fullplot}. The top half of the diagram shows collapsing universes, and the bottom half expanding universes. There are two saddle points (Einstein static universes $E1$ and $E2$) and a centre $F1$. The direction of flow means that, except for the separatrix that ends up at $E2$, for $\Omega>1$, a universe in the matter dominated epoch makes a transition from an expanding universe to a contracting one if it reaches $\Omega = \infty$. This occurs when the Hubble parameter is zero ($\dot{a}=0$) and so the density parameter $\Omega$ is infinite (this is just a coordinate singularity because of the representation (\ref{eq:Omega})).  If it does not reach $\Omega = \infty$, it expands forever with $\Omega$ asymptoting to unity, as do all models with $\Omega <1$. The line $\Omega=1$ is an attractor during the cosmological constant dominated epoch, as it was during inflation.\\

The source on the left-hand side is an initial radiative  Einstein-de Sitter universe, and the sink on the right-hand side is a final de Sitter universe. This diagram shows the symmetry of closed matter dominated universes that cycle as they expand, reach a maximum, and then collapse; however these do not correspond to the observable universe. There are three fixed points: two Einstein static universe $E1$, $E2$, which are solutions of the field equations, and the centre $F2$, which is not: it is due to the change of equation of state between matter dominated and a cosmological constant. There are   closed cycles around $F1$. The red solution beginning at the Einstein static universe $E1$ corresponds to the emergent universe  scenario  \cite{EllMaa04,Ellis:2003qz,Guendelman:2015qta}, where the universe starts off asymptotically as an Einstein static universe, passes through a phase like what we see today, and ultimately tends towards a flat universe, expanding forever due to the ultimate dominance of the cosmological constant. These can correspond to the universe today. \\

We note that there are models that exist for any measured $\Omega_{\rm Total}$ at the present day. The red bar represents our current point in the evolution of the universe, according to present observations.  It includes an emergent universe but no cyclic models (those that occur do not have a late time cosmological constant dominated era). 

\subsection{Phase planes with a decaying cosmological constant}
\label{decay}
 
So far we have studied a relatively standard picture. Now we consider dynamic dark energy, equivalent to a decaying cosmological constant. Again we use $\Omega=\Omega_m+\Omega_\Lambda$ to represent matter plus an effective cosmological constant. We model the latter as dynamic, decaying away as
\begin{equation}\label{eq:decay}
\Lambda(a) = \Lambda_0e^{-Ba},
\end{equation}
where $a$ is the scale factor and $B$ is a constant. 

\begin{figure}[H]
	\centering
	\includegraphics[trim = 1cm 8.5cm 2cm 8.2cm,scale=0.5]{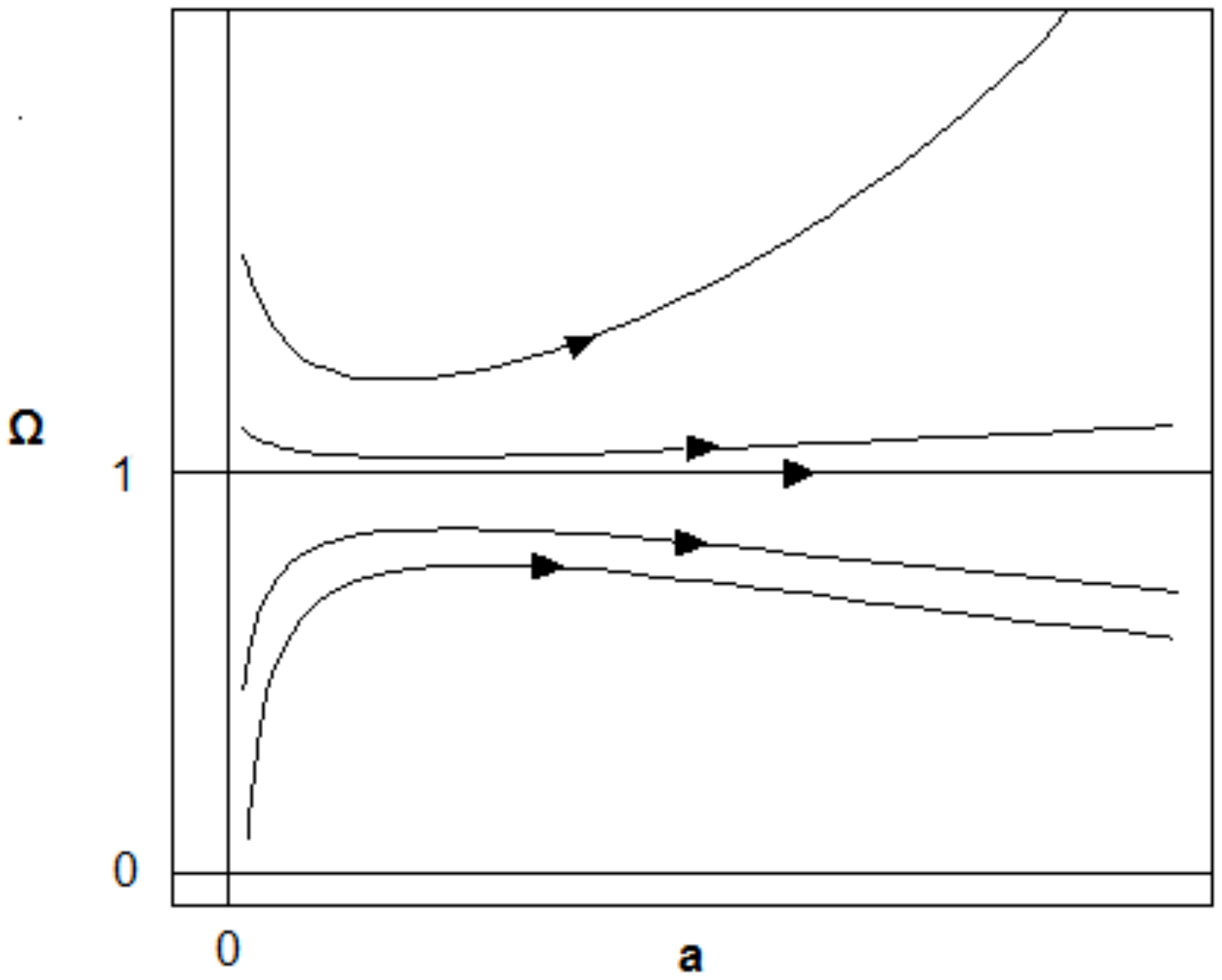}
	\caption{Phase plane for $\Omega=\Omega_m+\Omega_\Lambda$ with $\Lambda$ decaying}
	\label{lamdec}
\end{figure}
 
This system has an effective initial equation of state given by 
\begin{equation} \label{decaylambdaw}
w=\frac{Ba}{3}-1
\end{equation}
and hence for sufficiently small choice of $B$ begins with $w \approx -1$, evolving to $w=0$ as $\Lambda$ decays and matter begins to dominate, at $a=3/B$. Beyond $a=6/B$ the effective equation of state would have $w>1$, and such a model would be  invalid. In section \ref{discussion} we discuss a more physically motivated model of decay of the cosmological constant, which we can model as transferring energy into dust during its evolution. Its phase diagram is shown in FIG. \ref{lamdec}. The curves diverge away from $\Omega = 1$.

\subsection{Time symmetric Universe with decaying cosmological constant}

\begin{figure}[h]
 \centering
	\includegraphics[scale=0.6]{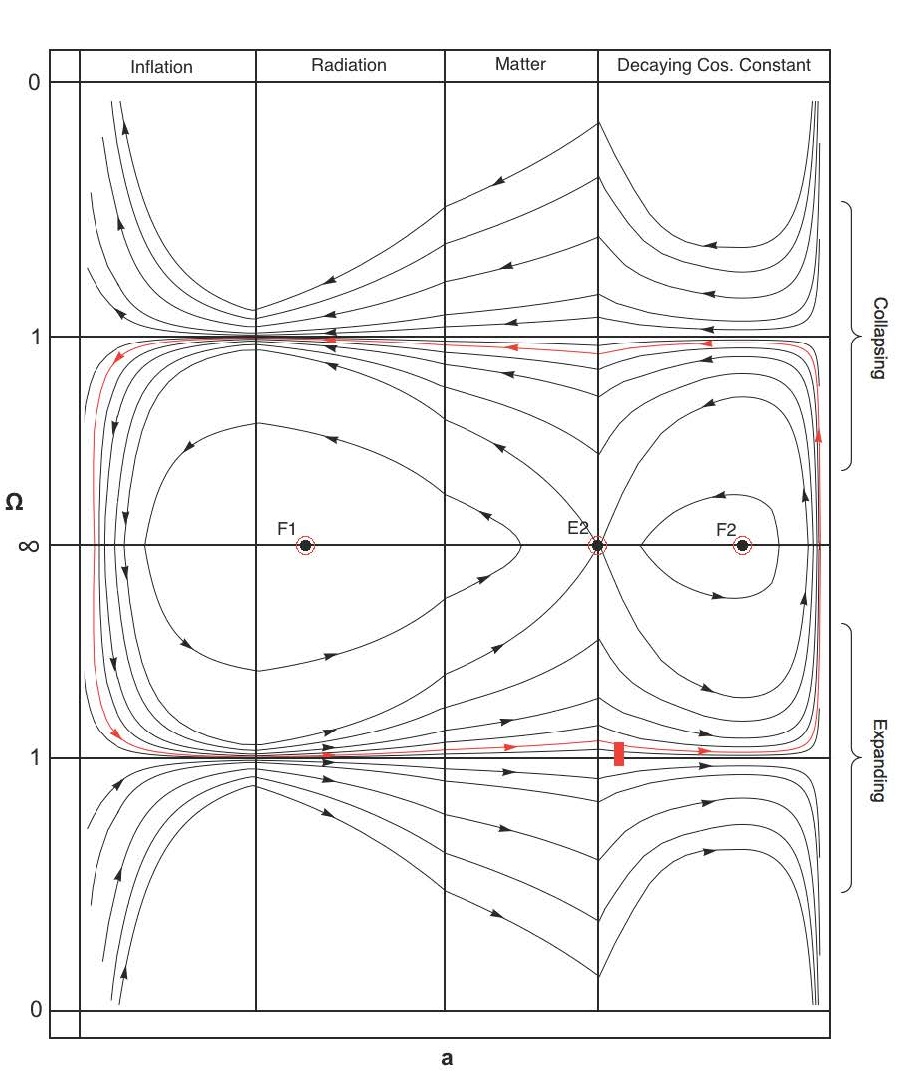}
	\caption{Time symmetric phase plane with a decaying cosmological constant}
	\label{fulldecay}
\end{figure}

Now we consider a situation like in section (\ref{sectcos}), but with a decaying $\Lambda$ of the form (\ref{eq:decay}) at late times. Again we create a time symmetric portrait (FIG. \ref{fulldecay}). We  assume here the universe is inflationary from the start, and so this time  exclude a  pre-inflation radiation dominated epoch. The results become quite interesting. 
\\

In the cosmological constant dominated era at late times, $\Lambda$ drives open and closed universes towards a flat universe as usually expected. However, after some point the cosmological constant has decayed away sufficiently that matter dominates again. This causes models with positive spatial curvature ($\Omega>1$) to slow down and then recollapse. The density parameter $\Omega$ of those models increases and diverges when $H=0$ for finite values of the scale factor, and this leads to the emergence of cyclic universes. A closed universe will cycle through phases of expansion and collapse, and an open universe will expand forever, tending towards an empty universe. Again we note that there are models that exist for any measured $\Omega_{\rm Total}>0$ at the present day. \\

The fixed points are indicated by red-circled black dots: two centres $F1$, $F2$, and an Einstein static universe $E2$.  
The fixed points $F1$ and $F2$ have closed cycles around them, while the Einstein static universe $E2$ is a saddle point.  We can calculate the radii of curvature for the Einstein static universes in FIG. \ref{fullplot} and FIG. \ref{fulldecay}. By setting $H=0$ in Equations \ref{cosmoone} and \ref{cosmotwo}, we obtain
\begin{equation}
R_E=\Lambda_E^{-1/2}=\frac{1}{\sqrt{4\pi G\rho}}.
\end{equation}
Using estimates for the inflationary potential energy and the current value of the cosmological constant of $\Lambda_{inf}\approx10^{14} GeV^4$ and $\Lambda_{DE}\approx10^{-12} GeV^4$, the radii of curvature are $R_{E1} \approx10^{-7} GeV^{-1}$ and $R_{E2} \approx10^{6} GeV^{-1}$. Thus if we modified this model to include a radiation dominated first stage, the first Einstein static universe (the source in the emergent universe scenario) would be much smaller than the second one (the static universe corresponding to the current value of dark matter). \\

The cyclic universes around $F1$ and $F2$ are not viable universe models, as they exclude the formative phases of the universe, such as inflation, reheating and recombination. However the red curve shows a cyclic universe that exhibits a history like that of our universe after its collapse phase and bounce at $\Omega= \infty$. \\

Note that there is a cyclic model for every value of $\Omega_0:=\Omega(t_0) > 1$ in FIG. 5 (where $t_0$ is the present time), however only those satisfying (\ref{eq:om1}), (\ref{eq:om2}) will be compatible with current observations. If the red bar represents our current position in the evolution of the universe,  one cyclic  universe that agrees with current observations of $\Omega_{\rm Total}$ is shown by the red lines in FIG. \ref{fulldecay}.  

\subsection{Time asymmetric Universe with decaying cosmological constant}

\textbf{Symmetry of cyclic phases} The above has assumed a symmetry between the collapse and expansion phases, but this is physically unrealistic. The exact cyclic models shown in the phase planes would in fact be replaced by progressively different cycles, as indicated first in the work of  
Tolman (\cite{Tolman1934}). He introduced the presence of a viscous fluid in an attempt to describe a cyclic universe, finding that each successive cyclic phase will have progressively shorter periods and larger amplitudes because of entropy increase (\cite{Tolman1934}:439-444). Barrow and Dabrowski \cite{Barrow:1995cfa} found an end to Tolman's growing cycles when introducing $\Lambda>0$, and for any $\rho+3p<0$. Rees \cite{Rees:1969} showed astrophysical considerations would make the collapse phases differ from expansion phases, \textit{inter alia} because of the different relations between stars and the radiative environment.\\

It has since been found, however, that a system with time-symmetric and non-dissipative governing equations can too possess an arrow of time \cite{Sahni:2012er}. This is through cosmological hysteresis, where an asymmetry in the equation of state during expansion and collapse causes an increase in amplitude of successive expansion cycles \cite{Sahni:2015kga}. The reversibility of the Hot Big Bang era presents a possible entropy problem, however this can be solved either through the introduction of new physics at the bounce, or by assuming a universal net baryon number of zero \cite{Rees:1969}. \\

Here,  we focus on the fact that the inflationary dynamics will not be time symmetric between the expansion and collapse phases, as assumed in the phase plane analysis above, because of the nature of inflationary dynamics. 

\section{Inflationary dynamics}
 
We assume that the inflationary expansion phase is that of a slow-roll field, as usual. It is the collapse phase that needs investigation: can it also be slow rolling? If so, will it remain reasonably spatially homogeneous so that we can continue to use a Robertson-Walker model for the geometry? We look at these issues in turn.

\subsection{Inflationary fields}
 	
The Klein-Gordon equation for a spatially homogeneous field in a FLRW universe is  
	\begin{equation}\label{eq:KG1}
	\ddot{\phi} + 3 H \dot{\phi} = - V'(\phi) ,
	\end{equation}
the energy density is given
\begin{equation}
\ \rho=\dot{\phi}^2/2 + V(\phi),
\end{equation}
and the pressure 
\begin{equation}
p = \dot{\phi}^2/2 - V(\phi) ,
\end{equation}
so we find that $w$ is given by
\begin{equation} \label{Infw}
w = \f{p}{\rho} =  \f{\dot{\phi}^2/2 - V(\phi)}{\dot{\phi}^2/2+V(\phi)}.
\end{equation}	
 
Consider two different simple cases for the collapsing phase in our $k=+1$ models.
	
\subsubsection{Flat potential}\label{sec:flat}
 
Suppose we have a flat potential:
	\begin{equation}
	\frac{\partial V (\phi)}{\partial \phi}=0.
	\end{equation}
	\begin{figure}[h]
		\label{Fig_asym}
		\centering
		\includegraphics[width=0.7\linewidth]{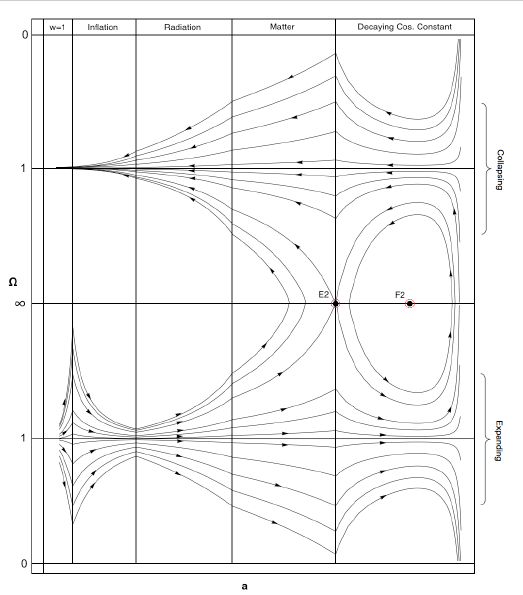}
		\caption{ Phase plane with $p = \rho$ in collapse phase.}
	\end{figure}
Then (\ref{eq:KG1}) becomes  	
	
	\begin{equation}
	\ddot{\phi} = -  3 \frac{\dot{a}}{a} \dot{\phi} 
	\Leftrightarrow	\frac{d \log \dot{\phi}}{d t} = - 3 \frac{d\log a}{dt}.
	\end{equation}
	Integrate
	\begin{equation}
	\log (\dot{\phi}/\dot{\phi_0}) = - 3 (\log (a/a_0)) 
\Rightarrow	\dot{\phi}/\dot{\phi_0} = (a/a_0)^{- 3} .
	\end{equation}
Hence as $a \rightarrow 0$, $\dot{\phi} \rightarrow \infty$ and the solution becomes velocity dominated:
	\begin{equation}
	\{a \rightarrow 0\} \Rightarrow \{p \rightarrow \rho > 0\} \Rightarrow \{\rho + 3p  > 0\}.
	\end{equation} 
A singularity occurs. We represent it by having the phase plane as before,  except the collapsing inflaton dominated phase now has equation of state 
	\begin{equation}
	p = \rho, \,\,\rho >0.
	\end{equation}
The solution is time asymmetric with no cyclic models except those around $F2$, which cannot represent the observed universe (FIG. 6).\\

All the observationally viable $k=+1$ models that reach a maximum and start contracting, collapse into a final singularity where $\Omega \rightarrow 1$ at a finite time in the future.

\subsubsection{Quadratic potential}\label{sec:quad}
 
Next we consider a quadratic potential
\footnote{Unlike the BICEP bounds that are now discredited, the current WMAP bound weakly excludes $\phi^2$ potentials. However that result depends on structure growth dynamics that might be altered in our models.}:   
	\begin{equation}
	V(\phi) = \frac{m^2}{2}\phi^2.
	\end{equation}
	Equation (\ref{eq:KG1}) becomes
	\begin{equation}\label{eq:KG3}
	\ddot{\phi} = -  3 \frac{\dot{a}}{a} \dot{\phi}  - m^2 \phi.
	\end{equation}
 	
Our current analysis encompasses only the homogeneous case, a fact which will be relaxed through the introduction of inhomogeneous perturbations in the next section. The precise nature of the dynamics of the inflaton field is highly sensitive not only to the potential, but also the initial conditions. As we are in a collapsing phase, Hubble parameter, $H$, is negative. Were it constant, we would be describing a damped harmonic oscillator with an anti-frictional term, the solutions of which are Bessel functions of the first and second kind. However, the Friedmann equation (\ref{freidone}) relates the Hubble to the energy density of the field and thus our system is more complicated than that described by a harmonic oscillator. Due to the time reversal invariance of the underlying physical model, we can describe the set of collapsing cosmologies as the reversal of the corresponding set of expanding solutions. Hence we find that there are solutions which exhibit all possible realisations of $w$ between $-1$ and $1$ over the course of collapse. There are solutions which will exhibit strong potential domination, strong velocity domination and all cases in-between during collapse. Particularly, there will be solutions which have $w<-1/3$ at the point where $\frac{8 \pi G \rho}{3} = \f{1}{a^2}$ and these will undergo a bounce (see section \ref{infbounce} ).\\

The result depends on the state of the inflaton after its rest energy has been exceeded in the collapse phase. In these models, we assume that the inflationary expansion phase is that of a slow-roll field, as usual. Provided the field is spatially homogeneous: $\phi = \phi(t)$, the collapse phase can also be slow rolling, and cyclic models are possible (for details, see the following sections).
	
\section{Chaotic cyclic cosmology}
 
So far the analysis has been based on the Robertson-Walker geometry; but the universe has small fluctuations about that geometry. The key further issue is if it will remain reasonably spatially homogeneous during the collapse phase, so that we can continue to use a Robertson-Walker model for the geometry.

\subsection{Inhomogeneity growth}
 
The growth of perturbations and matching perturbations at each cycle is a potentially problematic physical issue: one might expect inhomogeneities to grow during the collapse phase, and so destroy the Robertson-Walker like state of the universe \cite{Penrose:2008}. However this is not necessarily the case when one takes the inflationary equation of state into account.\\

To investigate this, it is convenient to use the 3+1 gauge invariant and covariant formalism \cite{Ellis:1989jt, Ellis:1989ju}, which centres on the comoving fractional spatial density gradient, defined as 
\begin{equation}
{\cal D}_a:= h_a^b \rho_{,b}/\rho
\end{equation}
for an observer with 4-velocity $u^a$ ($u_au^a = -1$), where $h_{ab}:=g_{ab}+u_au_b$ projects orthogonal to $u^a$  \cite{Ehlers:1993gf,Ell71}. When $w = p/\rho = const$, $\Lambda = 0$, and for any spatial curvature, the linearised growth equation for modes of wave number $n$ obtained this way is equation (38) in \cite{Ellis:1989ju}.  On using the Friedmann equation, it is
\begin{equation}
\label{de1}
\ddot{{\cal D}}_a  = - (2-3w)H\,\dot{{\cal D}}_a +\left(\frac{(1-w)(1+3w)}{2}
\kappa\rho\right) {\cal D}_a + w \frac{n^2}{a^2}{\cal D}_a.
\end{equation}

The signs of the terms on the right hand side of this equation depend on the equation of state as follows:
\begin{center}
\begin{table}[h]
	\begin{tabular}{|c|c|c|c|}
	\hline \rule[-2ex]{0pt}{5.5ex} Source & 1st term & 2nd term & 3rd term\\ 
		\hline \rule[-2ex]{0pt}{5.5ex} matter: $p=0: \,w = 0$ & $-2H$ &   $\frac{1}{2}\kappa\rho$ & $0$\\ 
		\hline \rule[-2ex]{0pt}{5.5ex} radiation: $p=\rho/3:\, w = 1/3$ & $-H$ &  $\frac{2}{3}\kappa\rho$& $\frac{1}{3}\frac{n^2}{a^2}$\\ 
		\hline \rule[-2ex]{0pt}{5.5ex} velocity dominated: $p=\rho:\, w = 1$ & $+H$ &  $0$& $\frac{n^2}{a^2}$\\ 
		\hline \rule[-2ex]{0pt}{5.5ex} potential dominated
		: $p=-\rho: \, w = -1$ & $-H$ & -$2\kappa\rho$ & $-\frac{n^2}{a^2}$ \\ 
		\hline 
	\end{tabular} 
	\caption{The terms in equation (\ref{de1}) for a variety of matter types. We see that the first term retards structure growth during expansion ($H>0$), acting as a friction force, but speeds it up during contraction ($H<0$). However the second term generates structure for ordinary matter and radiation, but causes oscillations for a cosmological constant, as does the third term.}
\label{termtable}
	\end{table}
\end{center}
 
During an inflationary collapse phase ($H < 0$), the first term soon dominates and is positive, and can make perturbations grow very fast, destroying the spatial homogeneity of the model:  the inhomogeneity catastrophe envisaged by Penrose when collapse takes place \cite{Penrose:2011zz} may occur. Whether this occurs or not depends on the initial conditions. In the next section we see that generically we may expect the result to be chaotic cyclic models. \\

Note that there are related calculations on perturbation growth in some other cyclic models, e.g.  \cite{Steinhardt:2001st, Khoury:2001wf, Khoury:2003rt, Turok:2004yx}, but they do not necessarily apply to our models as we have a singularity free bounce, in contrast to the other models studied. Calculations of perturbations in cyclic models are given in \cite{Piao:2009ku} in which it is found that universes undergoing such bounces will separate into causally disconnected regions. In the following section, we will see that such a scenario can be realised with some matter configurations. An example of a modified cosmology undergoing bounces in which a scalar field transitions between AdS vacua is given in \cite{Garriga:2013cix}, which comes to similar conclusions.

\subsection{Chaotic cyclic models}
 
Now we may be plausibly claim that what occurs is essentially the same as in the case of Linde's chaotic inflation  \cite{Linde:1986fc,Linde:1983gd,Brandenberger:1990wu}, but now occurring during the collapse phase.  During that phase, $\dot{a}<0$, we would expect that fluctuations of the inflaton would make it so that in some regions one finds $\dot \phi_0 >0 $ after the inflaton turns on, while in others $\dot \phi_0 <0 $. Then, as above, 
	\begin{itemize}
		\item \textbf{Case I} In some regions, we will get a velocity dominated collapse to a singularity, locally as in FIG. 6.
		\item \textbf{Case II} In some regions with $\dot \phi_0 >0 $ we may get slow rolling and  a potential dominated collapse phase;
		\item  \textbf{Case IIA} In some of those cases inhomogeneity will build up, and the Robertson-Walker geometry will no longer apply;  then the collapse phase will end up in many black holes;  
		\item \textbf{Case IIB} In the other regions, we will get a potential dominated  collapse to a minimum, followed by a bounce, locally as in FIG. 5.  
	\end{itemize} 
Hence one will get an inhomogeneous situation with piecewise singularities and piecewise bounces leading to a chaotic inflation type scenario with each cycle that leads to collapse producing a mix of singularities and bounces; and this can go  on for many bounces.  In many regions the universe will expand to a maximum and recollapse to a singularity, but in some it will collapse to a minimum and then re-expand; and because of the nature of inflationary expansion, it is the latter that will dominate the volume of the universe. Details are given in section \ref{infbounce}. 
	
\subsection{Quantum effects}
 
What happens if we take quantum fluctuations into account? In the case of standard chaotic inflation, these fluctuations can move the field up or down the potential. Presumably the same will hold in this case. This will tend to reinforce the chaotic cyclic picture presented here: some domains will get an even longer slow roll phase while others will get a more dominant velocity dominated phase. The quantum fluctuations, typically on the order of the Hubble parameter, may in some instances alter a velocity dominated collapse to one in which potential dominates and vice-versa. The overall picture will be qualitatively unchanged. The creation of a complete quantum model of such effects is worthy of further investigation, which we leave for future research. 

\section{Inflation Induced Bounces} \label{infbounce}
 
Let us consider the behaviour of a closed cosmology coupled to a scalar field with potential, as above (this is the model studied by Starobinsky in \cite{Starobinksy}). The Friedmann equation \ref{freidone} in the $k=1$ case allows for a vanishing of the Hubble parameter ($H=0$). This is usually associated with the recollapse of a large universe due to curvature effects, $H$, transitioning from positive to negative values. However, from Raychaudhuri's equation \ref{freidtwo} we see that if $w<-1/3$, then $\dot{H}>0$ --- a `bounce' from a collapsing universe to one which is expanding. This is of course well known at least in the case of de-Sitter universes, and corresponds to the scale factor following a hyperbolic cosine. However, in the case of a scalar field, $w$ is not a fixed parameter, but rather varies throughout evolution. Therefore it is entirely possible for a solution to experience both bounce and recollapse. Following equation (\ref{Infw}), $w=1$ for kinetic domination, $w=-1$ for potential domination, and so we will require that $V(\phi) > \dot{\phi}^2$ to have $w<-1/3$. If this is realised at the point where $H=0$, the bounce will occur. We can use this condition together with the Friedmann equation to find any choice of $w$ between $-1$ and $1$ can be achieved by setting 
\begin{equation}
 V(\phi) = \f{3(1-w)}{16 \pi a^2}, \quad\quad \dot{\phi}^2 = \f{3(1+w)}{8 \pi a^2}.
\end{equation} 
Thus strictly all we require is that $V(\phi)$ have a range which contains this value. For much of the further discussion, we will restrict to the quadratic potential commonly used for inflation, $V(\phi)= \f{m^2 \phi^2}{2}$, which for compatibility with observation sets $m^2=1.21 \times 10^{-12} m_{Planck}$. However, much of what is stated will be far more general. In this case, we find that the conditions that lead to a bounce give rise to an enormous amount of inflation. We are well into the region that will give rise to a prolonged period of slow-roll inflation, and hence will be compatible with observations.

\subsection{Background model bounce}
 
A question which arises at this point is whether one can find solutions that have such a behaviour. The answer is that such solutions do exist, but it can be hard to find them by doing a numerical search of the solution space at some point and then evolving forwards to the bounce. Most randomly chosen solutions at any given initial value of the Hubble parameter will not bounce. This is unsurprising, as the solutions which do go through the bounce are, moving away from the bounce, solutions which expand a lot in scale factor, and therefore attractors \cite{Corichi:2013kua,Sloan:2015bha}. Hence, if we run time backwards (or evolve towards a bounce in the future) we find that these solutions are repulsors --- if we aren't quite on the right solution we will quickly evolve away from it. Thus the way to find these solutions is to attack the problem from the other side: Specify conditions at the bounce (as is done above) and evolve away an expanding solution to some later time. Then reverse the direction of time on these solutions, and due to the temporal symmetry of the equations, these will be the solutions that describe collapsing solutions that undergo a bounce. These can be continued beyond the bounce and become expanding solutions on the other side.

\begin{figure}[htbp]
	\begin{center}
		\includegraphics[width=\linewidth]{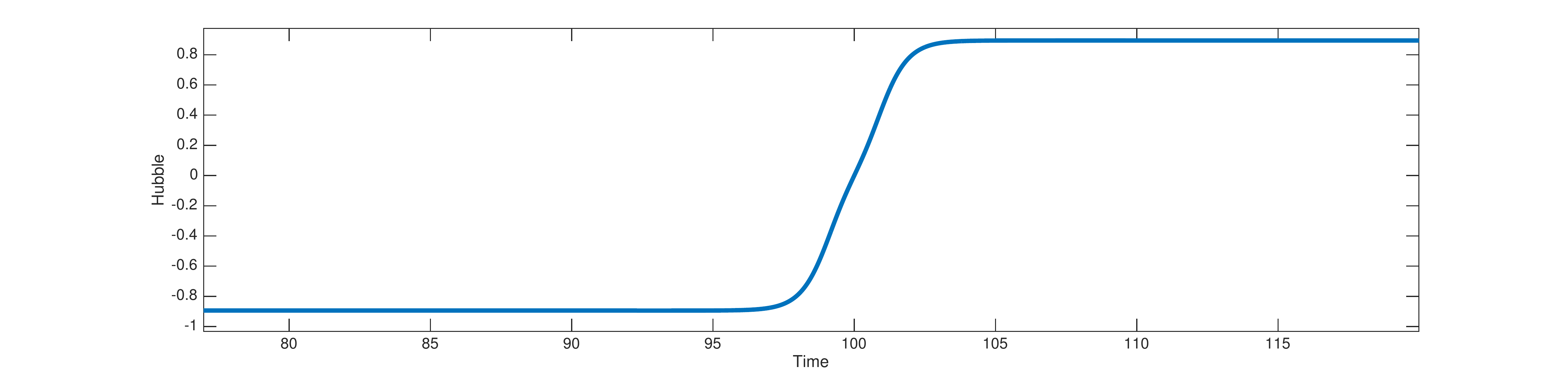}
		\caption{Hubble versus time for a bouncing solution for a limited time interval. Parameters used: $m=10^{-6}$, and at the bounce $w=-0.6$, $a=1$}
		\label{default}
	\end{center}
\end{figure}

\begin{figure}[htbp]
	\begin{center}
		\includegraphics[width=\linewidth]{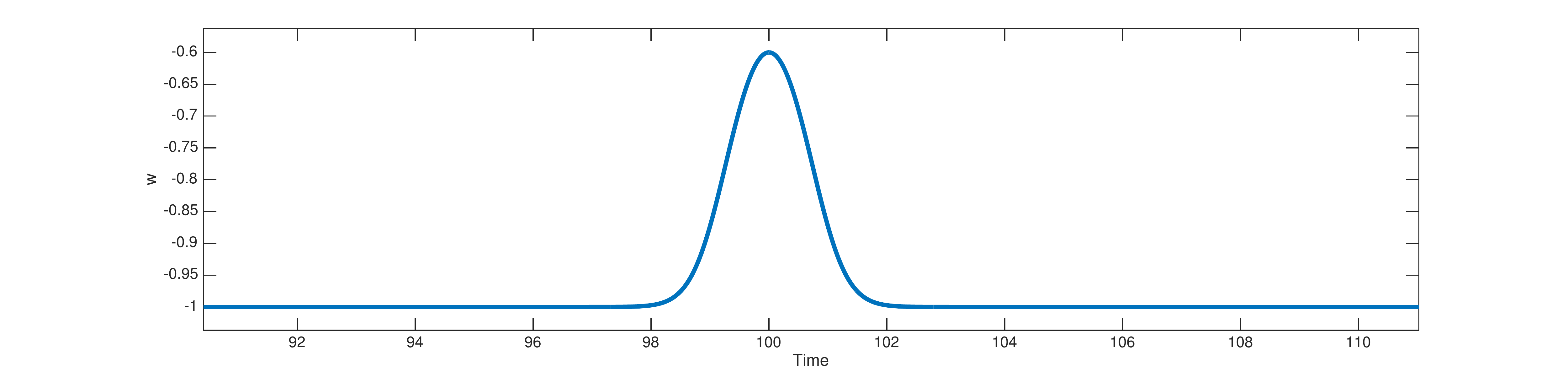}
		\caption{$w$ versus time for the above solution. We see that $w=-0.6$ at the bounce and quickly tends to $-1$, entering an extended period of slow-roll inflation.}
		\label{default}
	\end{center}
\end{figure}

\subsection{Perturbation growth}
 
One immediate problem that we encounter with such models is that the perturbations described in equation (\ref{de1}) grow in the contracting phase. This can be effectively seen by considering the second term in table (\ref{termtable}). We can qualitatively think about this as being a harmonic oscillator with a friction term. If we have a negative coefficient of the Hubble parameter there is friction in the expanding phase, but anti-friction in the contracting phase. This leads to an increase in the amplitude of the perturbations, and since there is a long period of slow roll, these perturbations can grow extremely large. 

\begin{figure}[htbp]
	\begin{center}
		\includegraphics[width=\linewidth]{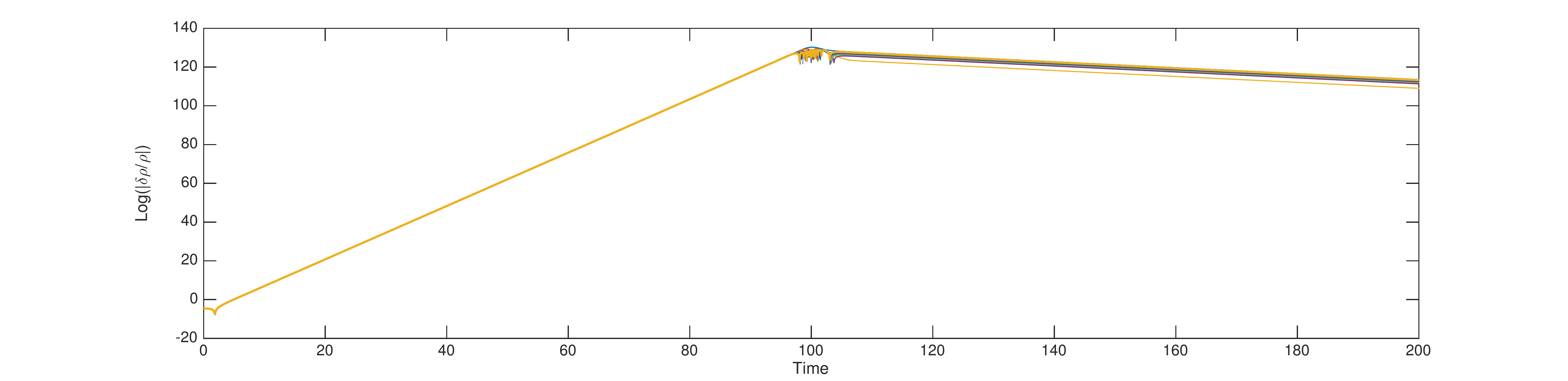}
		\caption{Perturbations grow exponentially with time in the contracting phase, then shrink again whilst expanding. Back reaction is ignored, and we see that from a .1\% perturbation in the energy density, and a relatively short period of slow roll, we overwhelm the background mode. Values: $w=-0.85$ at bounce, $\delta \rho = 0.001 \rho$ initially, quadratic potential with $m=10^{-3}$.}
		\label{default}
	\end{center}
\end{figure}
 
On first look it would appear that the growth of inhomogeneities would mean that the background model will be invalid as we approach the bounce. If we wish to avoid anti-friction during collapse, we would require $w>2/3$, which contradicts the requirements of the bounce ($w>-1/3$). However there are a few ways in which this can be resolved. We will look at two. 

\subsection{Special potentials} \label{Special}
 
The first possibility is that we choose an inflationary potential such that the universe is kinetic dominated for the most part, but there is a sudden and very steep rise. An example would be a quadratic for $\phi>0$ and no potential for $\phi<0$. In the kinetic region, $w=1$ and the amplitude of perturbations will be suppressed greatly. As we enter the potential dominated region, the potential starts to dominate, and $w$ gets larger. Amplitudes will start growing, but if the process is sufficiently fast (and it can be very fast as $\phi$ has been accelerated by a long period of anti-friction) the amplitudes won't grow by many orders of magnitude before the universe bounces. As such we hit the bounce with small amplitudes, and a high potential, friction takes over from anti-friction and we are returned an expanding universe undergoing slow-roll inflation. A (rather extreme) example of how this can be achieved is to consider a potential consisting of a well with vertical walls, of a fixed height. During collapse, whilst the energy density is sufficiently low, the scalar field will oscillate back and forth between the walls, gaining kinetic energy through the anti-friction mechanism. During this phase, the potential term is identically zero, and hence $w=1$, and so perturbation modes will be suppressed. However, once the energy density reaches a sufficient value, the inflaton will escape the well. If the walls are sufficiently close, this will transfer almost all kinetic energy to potential, pushing $w$ close to $-1$. This will be a sharp transition, and perturbations will once again begin to grow. However, if this happens sufficiently close to the bounce, the amplitude of perturbations will not have time to grow beyond their initial value and thus the homogeneous model remains valid and undergoes a bounce.  This mimics in some ways the behaviour of tunnelling between false vacua. 

\subsection{Filtering systems}
 
The second possible resolution is that across all of space there will be points in which there are large perturbations, but also there will probabilistically be regions in which these are small. To make this more precise, consider the past light-cone of a point that would see a bounce in the homogeneous case. Then we do not care about perturbations with wavelengths much longer than the radius of this lightcone at any given time, as these can be absorbed into the homogeneous background mode. Likewise, if we let the wave-number $n$ in equation $V.10$ become arbitrarily large, the oscillations will happen arbitrarily quickly, therefore the time-average of such perturbations over any given short time-span (such as a planck time) would be zero. Hence we only need a finite number of perturbative modes to be small in a given region, so the probability that such an event occurs is non-zero. Such regions, if they satisfy the condition of $w<-1/3$ will undergo a bounce. Regions not satisfying this will be forced into a singularity by the dynamics of their inhomogeneous modes (which contribute an effective term like $1/a^6$ and so dominate at small scales and kill the bounce effect). Therefore we will filter out large amplitude perturbation modes through such a bounce. \\

This is particularly neat as a solution, as it gives a natural explanation for what's going on that doesn't require anything to be hand-picked: We begin with a large, closed universe which is collapsing. During collapse the inflaton is formed with perturbations across space. Most regions are such that they will lead to a singularity, but some have the right conditions (potential domination, small fluctuations) for a bounce to occur. These conditions then filter space - most of the volume of pre-collapse space is in singularities, but some has survived and is now in the right configuration to undergo a long period of slow-roll inflation - a broadly homogeneous universe with small amplitude fluctuations. This inflation takes a small region and greatly expands it --- if we use favoured inflaton potentials, we get several orders of magnitude more e-folds than are required to match observations. This effectively pushes the singular regions far away from the inflated region, so what should be primordial singularities (black holes) in the expanding space-time will be washed away, and we are returned to a large, closed space. Thus we also have a natural explanation for the resetting of a local entropy: In a collapsing universe a large number of black holes will be formed, and these act as entropy sinks. It is only the highly ordered (low entropy) areas that even undergo a bounce, and hence the past horizon of a current observer will see lower entropy. Note that this is completely in keeping with the second law, as our system is not closed locally, only globally, and globally entropy is increasing as more black holes are formed, and the majority of a collapsing space ends up in one. Locally entropy is low after a bounce, precisely because that is a requirement of the filtering.\\

\begin{figure}[htbp]
	\begin{center}
		\includegraphics[width=\linewidth]{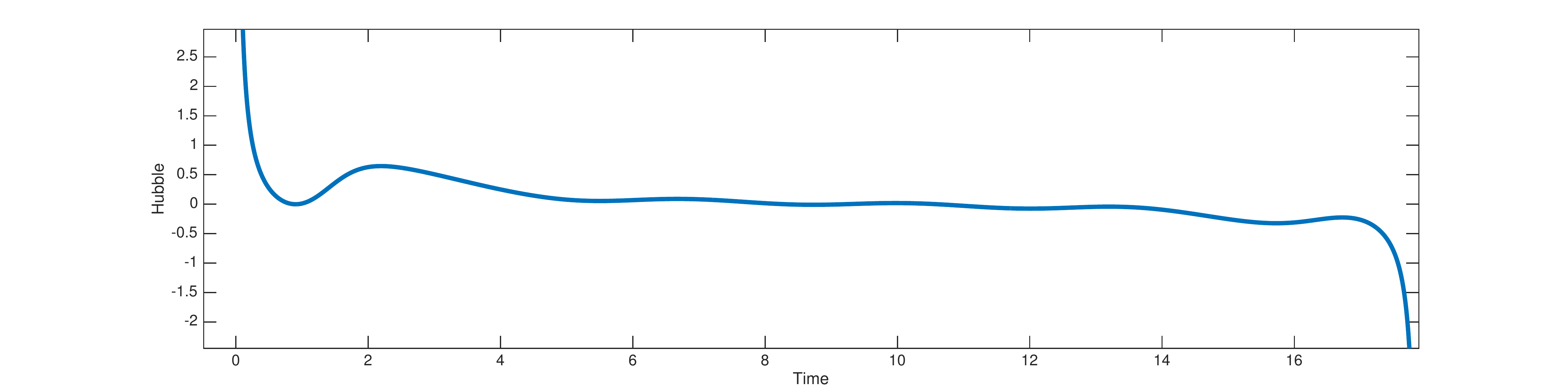}
		\caption{Hubble against time for a solution with a bounce (at $t=0.94$). We see a typical evolution with singularities in both the future and past, through bounces between them. This displays the behaviour of a universe that is initially expanding, collapses, bounces, collapses again, bounces again, and finally collapses down to a singularity. This is a toy model with $m=1$ and $w=-0.4$ at the first bounce.}
		\label{default}
	\end{center}
\end{figure}

\begin{figure}[htbp]
	\begin{center}
		\includegraphics[width=\linewidth]{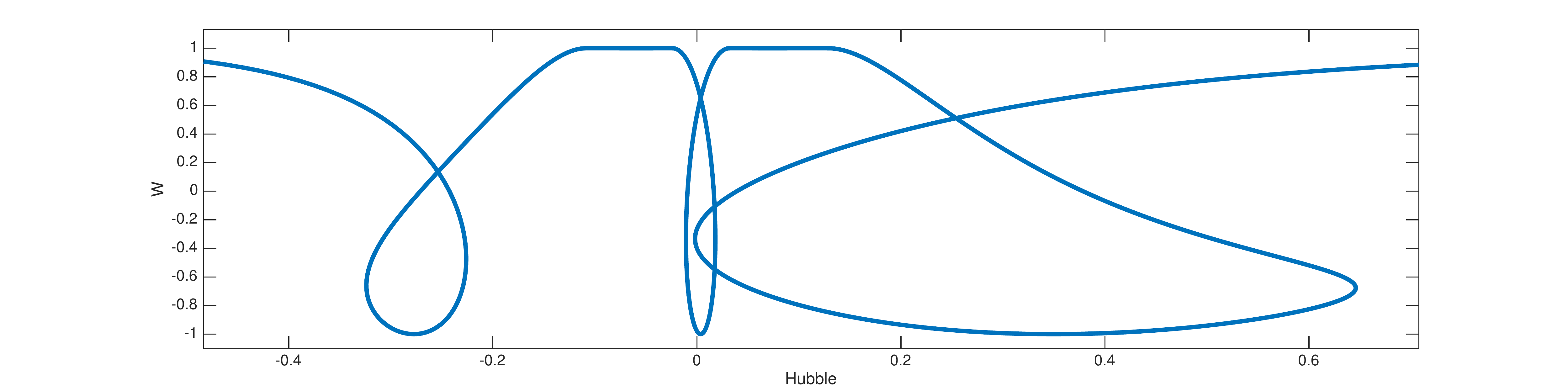}
		\caption{$w$ plotted against Hubble for the above system. The solution begins at the top right, and is expanding, undergoes recollapse with $w=-0.3$, bounces with $w=-0.4$, expands and recollapses with $w=0.6$, bounces again with $w=-1$ and recollapses a final time with $w=0.6$ again. This illustrates that multiple bouncing solutions with the same matter can occur.}
		\label{default}
	\end{center}
\end{figure}
 
In a typical quadratic system which matches with observations, the mass of the inflaton is roughly $10^{-6}$ in Planck units. A very approximate lower bound for the number of e-foldings can be obtained by making the approximation that slow roll will begin at approximately the potential at the bounce and will go on until the inflaton has a magnitude of order unity. As such, we find that there will be around 
\begin{equation}
N_e = Ht \sim \f{1-w}{2 \pi m^2} 
\end{equation}
e-folds between the onset and end of inflation in such a system. Hence we find that during inflation in such systems, the radius of the universe will increase by a factor of around $\exp(10^{12})$. Hence, we can arrive at an interesting estimate: If the probability of a pre-bounce patch having sufficiently small perturbation modes in the right set of wavelengths as defined above is more than $\exp(-10^{12})$, there will be more of such patches after inflation than before, and the cycle can continue indefinitely. Starobinsky estimates the probability of a bounce given the currently observed universe at $10^{-43}$ \cite{Starobinksy}, but this estimate does not include any conditions on the perturbation modes. Our system will naturally favour regions that undergo a minimal amount of deflation as these are the regions with the lowest number of wave modes that will fit in the past light-cone of a bouncing point. Hence we expect to see universes that collapse in almost slow-roll and cross $w=-1/3$ very close to the bounce. These can undergo as little as 1-2 e-folds of deflation, and so are likely not to have perturbations dominate.

\section{Compatibility with current data}
 
Now we need to relate these models to the data given at that start.

\subsection{Curvature}
 
$\Omega_k > 0$ is possible, see equations (\ref{eq:om1}), (\ref{eq:om2}). 
This is an absolute requirement for these models to be viable. If we ever prove $\Omega_k < 0$, they are done for. However it may be that $\Omega_k$ is so close to zero we will never determine its sign. The best unambiguous constraints on curvature will come from combining the radial and angular diameter distances measured in the baryon acoustic oscillation (BAO) experiments \cite{Takada:2015mma}. Several future experiments, planned to measure baryon acoustic oscillations, should reach these limits, including the Square Kilometer Array (SKA) project \cite{Bull:2015esa}, the Canadian Hydrogen Intensity Mapping Experiment (CHIME) \cite{chime}, and the Hydrogen Intensity and Real-time Analysis eXperiment (HIRAX) \cite{sievers}. Cosmic variance limits the dark energy independent best constraints to be $ \sigma(\Omega_K) \simeq 10^{-3}$. Specifically the forecasts from \cite{Takada:2015mma} predict an accuracy of the geometrical determination of the curvature to be $\sigma(\Omega_K) \sim 0.006$. BAO measurements could provide a full order of magnitude improvement in accuracy if we assume a model of dark energy. In particular, for $\Lambda$CDM the cosmic variance limited accuracy could go to $\sigma(\Omega_K) \simeq 2 \times 10^{-4}$ and for a simple model of dark energy parameterised by two equation of state parameters by $w(a) = w_0 + w_a(1-a)$, we can anticipate similar constraints on the order of a few $\times 10^{-4}$. Since BAO measurements determine curvature and dark energy together, this degeneracy is far harder to break if the dark energy model is more complicated with additional parameters. In either case, we are limited in our future accuracy in constraining curvature, but we may yet get a positive signature of curvature at the $10^{-3} - 10^{-4}$ level which would be very interesting, not only for this work, but for all early universe theories.   

\subsection{Dark energy decay}
 
Decay of dark energy at the present time is possible. Regarding it as a non-interacting perfect fluid, the energy conservation is  
\begin{equation}
\frac{d\rho_\Lambda}{dt} = -3 (\rho_\Lambda+p_\Lambda) H = -3 \rho_\Lambda(1+w) H,
\label{eq:rholam}
\end{equation}
so it will decrease in the future provided $w > -1$, which is compatible with (\ref{eq:de}). Note that furthermore it is perfectly possible, if perhaps unattractive, that the dark energy density is increasing at the present time but will decay at a later time. Indeed if we model it as  an effective scalar field, we can find a potential that will give virtually any behaviour \cite{ElMad91}, so phenomenological dark energy models that give an increase in the future followed by a decrease are possible. However what one would really like is a solid link to physics that will determine this evolution. No dark energy models at the present time provide this; however $w<-1$ creates great problems for both fluid and scalar field models, so a decreasing $\rho_\Lambda$ in the future is very plausible.  Whether it would decrease fast enough that matter dominates at some time in the future is the key issue for our models. An exponential decay, as envisaged here, will do the job; indeed any decay where $\rho_\Lambda$ decays faster than $1/a^3$ will also suffice, as then eventually baryonic matter $\rho_b$ will dominate over $\rho_\lambda$. \\

Assuming an exponential decay, a limit can be found on the parameters in equation \ref{eq:decay} from its derivative and with the current constraints on the equation of state (\ref{eq:de}). From this one obtains $B=-0.18^{+0.255}_{-0.273}$. Considering the upper limit, matter will dominate over dark energy at $a\approx48$, but the bounds, of course, allow for $B=0$ - a non-decaying constant.

\subsection{Inflationary constraints}
 
The models must fit the usual inflationary constraints, as given by Planck, WMAP, BICEP, and other projects. This will not cause problems for those models that bounce and then have a sufficient number of e-folds in the expanding inflationary era, as discussed in the preceding section. Although quadratic potentials are somewhat disfavoured compared to models which predict a smaller tensor to scalar ratio \cite{Ade:2015xua}, they provide a natural testing ground for the models discussed. In such models the inflaton mass is determined from the amplitude of fluctuations and the spectral index. Since the conditions for a bouncing universe require that the inflaton be potential dominated at the point of bounce, a long period of slow roll inflation will follow for any matter model which is currently favoured by observations. Data from the joint BICEP-Keck array of observations of the CMB place strong bounds on the scalar-tensor ratio, stating that $r<0.07$ at the 95\% confidence level, and $n_s\approx 0.97$ \cite{Array:2015xqh}. These bounds exclude the dynamics of a single scalar field inflaton, but are easily fit to quartic potential fields, or multi-field inflation \cite{Easther:2013rva,Price:2015qqb}. Our analysis is agnostic to such distinctions --- there will be quantitative differences between the dynamics of cosmology depending on these potentials, however since all such potentials allow for $w<-1/3$, any of them can lead to a bounce. 

\section{Discussion} \label{discussion}
 
 A key issue for cosmology today is whether the universe is cyclic or not. In this short article, we have shown is that it is possible, at least in principle, that the universe was not singular at the start of the present expansion phase. It may have been singular prior to the first inflationary stage, but it may be that we have been through many cycles before the current phase of acceleration. Most tantalising is the fact that all current observations are consistent with us living on the red curve representing a cyclic universe. Current and future constraints on $\Omega_k$ are not expected to improve beyond the $10^{-3} $ level \cite{Santos:2015gra, Takada:2015mma} and constraints on the evolution of the equation of state of dark energy are set at $\frac{dw}{da}=0.03^{+0.68}_{-0.40}$ \cite{Kumar:2012gr}. Of course, dark energy could also begin to evolve in a different way at some point in the near future, and we do not have any constraint on that, though a compelling physical mechanism would need to be found to make this result worth seriously considering. \\

\textbf{Singularity Theorems} Borde et al. have given claims that the universe must have been singular \cite{Borde:2001nh} (BGV), but these are based on restrictive assumptions that need not hold in the real universe. In particular, the BGV theorem relies on two assumptions; either there is only expansion, or the average expansion is positive $H_{av} >0$ \textit{along a specific geodesic}, and the second is that the behaviour can be modelled entirely classically. In some instances of our model we consider a universe that undergoes periods of expansion and collapse; there exist some geodesics along which the average expansion, $H_{av}=0$ or can be negative across a cycle. To see this, consider a small patch of space which undergoes a bounce at the bounce point. The past of this patch contains a period of deflation, and a period of inflation. In the case of the special potential mentioned in section \ref{Special}, the energy density of the inflaton is time symmetric, with the motion only sensitive to the relative phase of the inflaton field before the transition between potential and kinetic domination, and thus we would have an almost perfect symmetry between expansion and contraction --- $H_{av} \approx 0$. This can be negative if the phase is such that the inflaton falls down the potential well whilst still retaining some kinetic energy in the expanding phase, yet encounters the wall with very little left-over potential energy in the contraction phase, and the symmetric condition will of course yield $H_{av}=0$. The second condition is that the behaviour is entirely classical. For any of our inflating systems, during inflation there is a period in quantum fluctuations of the inflaton field are assumed to be converted into classical perturbations of the fields. The quantum fluctuations are assumed to arise from the vacuum spontaneously, a behaviour not modelled by classical dynamics. If one were to take the classical theory with the perturbation modes and time-reverse the motion, one would find that as the universe contracted, these modes would quickly become trans-planckian, and dominate the motion entirely leading to an entirely physically distinct behaviour from that which occurred in the past of the model. It is therefore not correct to invoke the BGV theorem in this context as the behaviour of a classical geodesic during this phase is not well established.  We therefore evade this theorem simply by not satisfying the required axioms for the theorem to hold true in either instance. There are further counterexamples to this theorem, namely emergent universes \cite{Ellis:2003qz,Guendelman:2015qta} because a scalar field can the violate energy condition (as in inflation). These have been criticised as not being stable to quantum perturbations \cite{Mithani:2011en, Mithani:2012ii, Mithani:2014toa, Mithani:2014jva} but they can be so with the correct potential \cite{delCampo:2010kf, delCampo:2015yfa}.  \\

\textbf{Other bounces} Various approaches to quantum gravity have produced models which undergo bounces. A prime example of this is Loop Quantum Cosmology (LQC) \cite{Ashtekar:2011ni,Ashtekar:2006wn} in which the geometrical variables are polymer quantised to mimic the holonomy-flux algebra of Loop Quantum Gravity. The qualitative dynamics of these models differs significantly from that which is discussed here: In loop cosmologies there are no singularities at all \cite{Singh:2009mz,Singh:2010qa}. Although slow roll inflation is highly likely in LQC \cite{Ashtekar:2011rm}, it is not inevitable in the bouncing models as we have found here. Similarly, models of ekpyrosis \cite{Steinhardt:2001st} introduce a new physical forces motivated by string/M theory to the Friedmann equations, which lead to the contracting phase being attached to an expanding branch. Similarly the introduction of exotic matter, such as ghosts and phantoms can lead to a bounce of a collapsing universe. Particularly interesting phenomenologically is the model of varying the fine structure constant introduced by Bekenstein \cite{Bekenstein:1982eu}, which lead to the Bekenstein-Sandvik-Barrow-Magueijo (BSBM) models of cosmology. In such models the varying of $\alpha$ leads to a negative contribution to the energy density which can overwhelm collapse and produce a vanishing of the Hubble parameter at finite scale factor \cite{Barrow:1998df}. Our models differ in a distinct manner from all those mentioned here, as we need invoke neither new physical regimes (such as quantum gravity) nor new types of matter. Rather we require only that there exists some matter whose equation of state allows $w<-1/3$, which we model after the inflaton. \\

 \textbf{Phase planes and existence of cyclic models} We have shown that the Friedmann equations can be used to create phase planes for the cosmological density parameter $\Omega$ with respect to the rate of expansion $a$. Different values of $w$ represent different epochs through which the universe has evolved. The time symmetric phase plane in \cite{Madsen:1992tv} was updated to include the dark energy dominated epoch considering two options. We completed the phase plane analysis for both a true and a decaying cosmological constant. In the former case, we found that the universe would expand forever, tending towards a flat universe. For a decaying cosmological constant, the behaviour becomes more interesting. Open universes expand forever, tending towards an empty universe, while closed universes become cyclic. This model is possible with the current constraints on cosmological parameters, in particular with $\Omega=1\pm0.034$ \cite{Ade:2015xua}. This analysis offers a good visual example of a cyclic universe that is in keeping with current observations. With a decaying cosmological constant, and positive spatial curvature, no special mechanisms are needed to invoke a cosmic turnaround. \\

 There are many subtleties inherent in the physics that are not captured by the Friedmann equations or even by linear perturbations. For example, it was also suggested by Saslaw \cite{Saslaw:1991} that any black holes present in a contracting phase of a closed universe may lead to an entropy `catastrophe' and that the resulting inhomogeneities can lead to re-expansion in a significant portion of the universe. He further conjectured that surviving black holes could form the seeds for new structure formation. A further criticism that may arise is that the nature of perturbations may overwhelm the contribution from curvature and $\Omega_k$ can be arbitrarily small. However this reasoning if erroneous: there is always spatial curvature within our model as this is determined by the topology of the spatial manifold, which we do not allow to change (see for reasons such a topology change is unlikely). Hence all we require is that $\Omega < 1$, no matter how small this difference may be. \\

 \textbf{The physics of the models} Unlike a variety of cyclic models on the market  that assume speculative physics involving extra dimensions or modifications of general relativity, the models presented here require no substantial new physics --- they require only that the current dark energy, which is verified to be there by many observations, decay in the future until its energy density is less than that of the matter.  It is common cause that dark energy may have a dynamic character, indeed numerous phenomenological models explore this possibility. The proposal we put forward is very conservative, as regards the physics involved. \\

There are a few ways in which a decaying $\Lambda$ of section \ref{decay} can be achieved: The most natural  is to introduce a constant to the scalar field potential used for the inflaton. This can be arbitrarily small to give the right value of $\Lambda$. Then one can couple such a field (or indeed any other field $\Lambda$) to any other matter present in the system \cite{Barrow:2013qfa}. If we introduce a coupling between different fluids, then one can alter their dynamics. For a set of coupled fluids, we simply add a transfer term to the evolution equations: 
\begin{equation}
\dot{\rho_i} + 3H(\rho_i + p_i ) = \sum_j \gamma_{ij} 
\end{equation}
for some antisymmetric $\gamma_{ij}$ which models energy flow between fluids. Energy conservation in matter is achieved since $\gamma$ is antisymmetric (what leaves one field enters another) and we can make $\gamma$ a function of whatever variable we find appropriate. So at late time one can introduce a $\gamma$ between cosmological constant and dust, say, and have $\Lambda$ bleed away in that manner. Likewise one could have the scalar field coupled to dust, and by introducing a $\gamma_{\Lambda d}$ which is a function of $H$ have this occur only at late times. A similar method can be used to model flow of energy out of the inflaton and into, say, radiation. A particularly slick way of doing this is to have $\gamma_{ij}$ contain a $\delta(w_1 - w_2)$. That way the inflaton will dump energy into radiation, dust, etc as it descends on its potential, and also give a nice mechanism for returning energy to form the inflaton on a collapsing phase. A numerical simulation of this is beyond the scope of the current paper, but would be an interesting model to look at, and realises a decaying cosmological constant in a rather natural manner.\\

Furthermore, it is possible that with further investigation the cosmological constant may in fact turn out to be a relic of early data surrounding Type $Ia$ supernovae \cite{Nielsen:2015pga}. Should the evidence for a late time accelerated expansion of the universe go away, one will obviously be able to get cyclic models of the kind envisaged above but without the need for a mechanism to make the cosmological constant decay. \\

Following the turn around, there is a time-reversed collapsing hot big bang phase after the matter ionisation energy is exceeded, and eventually the original inflaton will resurrect, as such, and cause a $k=+1$ inflationary epoch with $a(t) = \cosh Ht$ in those domains that re-expand. In this sense, it is the original inflaton itself that causes the bounce, hence no new physics is involved. \\

\textbf{The geometry of the models} Many papers on inflationary cosmology consider only $k=0$ models.  This paper shows how limiting that constraint is; and emphasizes how important observational constraints on the sign of spatial curvature are. Determining that sign should be a major aim of observational cosmology. \\

Many interesting questions remain regarding the physical viability of such a cyclic model given the usual problems associated with cyclic models, including the growth of entropy and the growth of perturbations through each cycle. The detailed physics of an inverse reheating process are also to be explored in detail, given that the physics of reheating itself may not be time symmetric. A further consideration would be how to construct an explicit decaying cosmological constant model that would fit with all available observations to date. We leave these interesting and open questions to future work.

\section{Acknowledgements}

The authors are indebted to the anonymous referee whose remarks have improved this work. We would like to thank Jeff Murugan and Jean-Philippe Uzan for very useful discussions, and particularly Pedro Ferreira for a key remark. GE thanks the Physics Department, Oxford University, for hospitality. This work is based on the research supported by the South African Research Chairs Initiative of the Department of Science and Technology and National Research Foundation of South Africa as well as the Competitive Programme for Rated Researchers (Grant Number 91552) (AW and EP). EP is also supported by a Masters Bursary from the South African National Institute for Theoretical Physics (NITheP). GFRE is supported by NRF grant 96031.  Any opinion, finding and conclusion or recommendation expressed in this material is that of the authors and the NRF does not accept any liability in this regard. 

\bibliographystyle{jcap}
\bibliography{phaseplanesfinalv3}

\newpage
\appendix
\section*{Appendix: Bounces in inflationary cosmology}

\subsection*{Notes on numerics}
 
There follow a few notes on how the setup is achieved and ways in which systematic errors (such as numerically missing solutions due to machine error) can be avoided. The code is in units wherein $\f{8 \pi G}{3} = 1$. This isn't simply for efficiency of coding, but also eliminates a lot of $\pi$ factors (and particularly their square roots) in equations --- the numerical approximations of which introduce small errors that build over time. Also, in order to avoid dealing with large numbers,  it is more efficient to evolve a system consisting of $\{H,\phi,\dot{\phi}\}$. Since the scale factor becomes arbitrarily large, it isn't a good thing to keep around --- again 32-bit precision isn't enough when $a$ can grow by 30-40 orders of magnitude during inflation! The dynamical system we want is therefore determined by evolving the coupled ODE's:

\begin{eqnarray}
\f{d}{dt} &\phi& = \dot{\phi}, \nonumber \\
\f{d}{dt} &\dot{\phi}& = -3 H \dot{\phi} - V'(\phi) \nonumber, \\
\f{d}{dt} &H& = V(\phi) - \dot{\phi}^2 - H^2. \end{eqnarray} 
 
This is a closed system, therefore we don't need the scale factor to evolve it. The curvature is introduced by the constraint on the initial data surface. In this system, we can define conditions at a bounce by setting:

\begin{eqnarray} V(\phi)&=& \f{1-w}{2a_i^2}, \nonumber \\
\f{\dot{\phi}^2}{2} &=& \f{1+w}{2a_i^2}, \nonumber \\
H &=& 0 .
\end{eqnarray}
for some initial value of the scale factor $a_i$ (which is irrelevant, hence can just  be set to be 1). So we can pick $w$ at the bounce (recall that this must be less that $-1/3$ for a bounce, otherwise we're defining a recollapse). \\

Once we have a solution to this system, we can obtain $a(t)$ by integrating $H$, and find $w(t)$ etc from kinetic and potential energies. Once we have all this data, we can numerically integrate the equations defining perturbations in terms of these variables which is again relatively straightforward. 

\end{document}